\documentstyle[12pt,epsf]{article}

\expandafter\ifx\csname amssym12.def\endcsname\relax \else\endinput\fi
\expandafter\edef\csname amssym12.def\endcsname{%
       \catcode`\noexpand\@=\the\catcode`\@\space}
\catcode`\@=11
\def\undefine#1{\let#1\undefined}
\def\newsymbol#1#2#3#4#5{\let\next@\relax
 \ifnum#2=\@ne\let\next@\msafam@\else
 \ifnum#2=\tw@\let\next@\msbfam@\fi\fi
 \mathchardef#1="#3\next@#4#5}
\def\mathhexbox@#1#2#3{\relax
 \ifmmode\mathpalette{}{\m@th\mathchar"#1#2#3}%
 \else\leavevmode\hbox{$\m@th\mathchar"#1#2#3$}\fi}
\def\hexnumber@#1{\ifcase#1 0\or 1\or 2\or 3\or 4\or 5\or 6\or 7\or 8\or
 9\or A\or B\or C\or D\or E\or F\fi}

\font\tenmsa=msam10 scaled\magstep1
\font\sevenmsa=msam7 scaled\magstep1
\font\fivemsa=msam5 scaled\magstep1
\newfam\msafam
\textfont\msafam=\tenmsa
\scriptfont\msafam=\sevenmsa
\scriptscriptfont\msafam=\fivemsa
\edef\msafam@{\hexnumber@\msafam}
\mathchardef\dabar@"0\msafam@39
\def\dashrightarrow{\mathrel{\dabar@\dabar@\mathchar"0\msafam@4B}}
\def\dashleftarrow{\mathrel{\mathchar"0\msafam@4C\dabar@\dabar@}}

\def\ulcorner{\delimiter"4\msafam@70\msafam@70 }
\def\urcorner{\delimiter"5\msafam@71\msafam@71 }
\def\llcorner{\delimiter"4\msafam@78\msafam@78 }
\def\lrcorner{\delimiter"5\msafam@79\msafam@79 }
\def\yen{{\mathhexbox@\msafam@55 }}
\def\checkmark{{\mathhexbox@\msafam@58 }}
\def\circledR{{\mathhexbox@\msafam@72 }}
\def\maltese{{\mathhexbox@\msafam@7A }}

\font\tenmsb=msbm10 scaled\magstep1
\font\sevenmsb=msbm7 scaled\magstep1
\font\fivemsb=msbm5 scaled\magstep1
\newfam\msbfam
\textfont\msbfam=\tenmsb
\scriptfont\msbfam=\sevenmsb
\scriptscriptfont\msbfam=\fivemsb
\edef\msbfam@{\hexnumber@\msbfam}
\def\Bbb#1{{\fam\msbfam\relax#1}}
\def\widehat#1{\setbox\z@\hbox{$\m@th#1$}%
 \ifdim\wd\z@>\tw@ em\mathaccent"0\msbfam@5B{#1}%
 \else\mathaccent"0362{#1}\fi}
\def\widetilde#1{\setbox\z@\hbox{$\m@th#1$}%
 \ifdim\wd\z@>\tw@ em\mathaccent"0\msbfam@5D{#1}%
 \else\mathaccent"0365{#1}\fi}
\font\teneufm=eufm10 scaled\magstep1
\font\seveneufm=eufm7 scaled\magstep1
\font\fiveeufm=eufm5 scaled\magstep1
\newfam\eufmfam
\textfont\eufmfam=\teneufm
\scriptfont\eufmfam=\seveneufm
\scriptscriptfont\eufmfam=\fiveeufm

\csname amssym.def\endcsname

\newif{\ifcomentarios}
\comentariosfalse
\renewcommand{\theequation}{\thesection.\arabic{equation}}
\newtheorem{theorem}{Theorem}
\newtheorem{lemma}[theorem]{Lemma}

\newtheorem{proposition}[theorem]{Proposition}

\newtheorem{remark}[theorem]{Remark}
\newtheorem{example}[theorem]{Example}

\newcommand{\zerarcounters}
{
\setcounter{equation}{0}
\setcounter{theorem}{0}
}

\newcommand{\calA}{{\cal A}}
\newcommand{\calB}{{\cal B}}
\newcommand{\calC}{{\cal C}}

\newcommand{\calH}{{\cal H}}

\newcommand{\calN}{{\cal N}}
\newcommand{\calO}{{\cal O}}

\newcommand{\calU}{{\cal U}}

\newcommand{\calX}{{\cal X}}

\newcommand{\be}{\begin{equation}}
\newcommand{\ee}{\end{equation}}
\newcommand{\bma}{\begin{displaymath}}
\newcommand{\ema}{\end{displaymath}}
\newcommand{\bc}{\begin{center}}
\newcommand{\ec}{\end{center}}

\newcommand{\text}{\rm}

\newcommand{\dfrac}{\displaystyle\frac}

\newcommand{\dlim}{\displaystyle\lim}

\newcommand{\QED}{\samepage\bigskip\hfill\Box}

\newcommand{\uflex}
{{\scriptstyle {\raise 9pt\hbox{$\backslash$}\,\!\!\!\!\!\Bigg\vert}}}

\newcommand{\Z}{\Bbb Z}

\newcommand{\R}{\Bbb R}
\newcommand{\C}{\Bbb C}

\newcommand{\UM}{{\it 1\! \! \! 1}}
\newcommand{\I}{{\rm  I}}

\newcommand{\ncm}{\newcommand}
\ncm{\rncm}{\renewcommand}
\ncm{\id}{{\bf 1}}
\ncm{\beq}{\begin{equation}}
\ncm{\eeq}{\end{equation}}
\ncm{\ba}{\begin{array}}
\ncm{\bea}{\begin{eqnarray}}
\ncm{\beanon}{\begin{eqnarray*}}
\ncm{\ea}{\end{array}}
\ncm{\eea}{\end{eqnarray}}
\ncm{\eeanon}{\end{eqnarray*}}
\ncm{\fns}{\footnotesize}
\ncm{\setc}[1]{\setcounter{equation}{#1}}
\newcounter{eqnr}
\newenvironment{eqnarrayabc}{\stepcounter{equation}
  \setcounter{eqnr}{\value{equation}}\setc{0}
  \rncm{\theequation}{\thesection.\arabic{eqnr}\alph{equation}}
  \begin{eqnarray}}{\end{eqnarray}\setc{\value{eqnr}}}
\ncm{\eqboxabc}[3]{\newline\parbox[t]{1.5cm}{#1}\hfill
  \parbox[b]{12cm}{\begin{eqnarray*} #3\end{eqnarray*}}\hfill
   \parbox[b]{1.5cm}{\vspace{-0.0cm}\begin{eqnarrayabc}#2\end{eqnarrayabc}}\newline}
\ncm{\eqbox}[2]{\newline\parbox{1.5cm}{#1}\hfill
  \parbox{12cm}{\beanon #2\eeanon}\hfill
  \parbox{1cm}{\bea\eea}\newline}
\ncm{\nr}[1]{\parbox{1cm}{\begin{eqnarrayabc}#1\end{eqnarrayabc}}\\}

\ncm{\kal}[1]{\mbox{$\cal #1 $}}
\ncm{\mrk}[1]{\!\!\! #1 \!\!\!} 
\ncm{\qed}{\hspace*{0.4cm}\rule{0.24cm}{0.24cm}}  
\ncm{\mbold}[1]{\mbox{\boldmath $ #1 $}}   
\ncm{\bm}{\mbold}
\ncm{\str}{\stackrel}
\ncm{\sub}{\subset}
\ncm{\e}{\varepsilon}
\ncm{\ka}{\kappa}
\ncm{\inputc}[1]{\begin{center}\input{#1}\end{center}}
\ncm{\lto}{\longrightarrow}
\ncm{\x}{\times}
\ncm{\bmm}{\bm{\cal M}}
\ncm{\cp}{{\bf P}}    
\ncm{\bfp}{{\bf P}}
\ncm{\bmi}{\bm{i}}
\ncm{\bmom}{\bm{\om}}
\ncm{\bmOm}{\bm{\Om}}
\ncm{\res}{\restriction}
\ncm{\bmL}{\bm{\cal L}}
\ncm{\bmell}{\bm{\ell}}
\ncm{\bmE}{\bm{\cal E}}
\ncm{\bme}{\bm{e}}
\ncm{\bmpi}{\bm{\pi}}
\ncm{\bmr}{\bm{r}}
\ncm{\bmsigma}{\bm{\sigma}}
\ncm{\wt}{\widetilde}

\newcommand{\beaa}{\begin{eqnarray}}
\newcommand{\eeaa}{\end{eqnarray}} 
\newcommand{\Proof}{\noindent  Proof.}

\begin{document}

\author{{\bf Oscar Bolina}\thanks{Supported by FAPESP under grant
97/14430-2. {\bf E-mail:} 
bolina@lobata.math.ucdavis.edu} \\
Department of Mathematics\\
University of California, Davis\\
Davis, CA 95616-8633, USA\\
\\
{\bf Domingos H. U. Marchetti}\thanks{Partially supported by
FAPESP under grant 95/0790-1 and CNPq. {\bf E-mail:} marchett@ime.usp.br}  \\ 
Instituto de F\'{\i}sica \\
Universidade de S\~ao Paulo\\
Caixa Postal 66318\\
05315-970 S\~ao Paulo, Brasil\\
}
\title{\vspace{-1in}
{\bf The Falicov-Kimball Model \\
with Long--Range Hopping Matrices}}
\date{}
\maketitle
\begin{abstract}
\noindent
The ground state nature of the Falicov-Kimball model 
with unconstrained hopping of electrons is investigated. We solve the
eigenvalue problem in a pedagogical manner and give a complete account of the
ground state energy both as a function of the number of electrons and nuclei
and as a function of the total number of particles for any value of
interaction $U\in \R $. We also study the energy 
gap and show the existence of a phase transition characterized by the
absence of gap at the half--filled band for $U<0$. 
The model in consideration was proposed and solved by Farkas\u
ovsky \cite{F} for finite lattices and repulsive on-site interaction $U>0$.
Contrary to his proposal we conveniently scale
the hopping matrix to guarantee the existence of the
thermodynamic limit. We also solve this model with bipartite unconstrained
hopping matrices in order to compare with the Kennedy--Lieb variational
analysis \cite{KL}.

\bigskip
\bigskip

\noindent
{\bf Key words:} Falicov-Kimball model; Long-range hopping; Ground
state; Phase Transition \hfill \break
\end{abstract}



\section{Introduction}
\zerarcounters

The Falicov-Kimball model \cite{FK} of correlated (spinless) electrons
on a lattice $\Lambda \subset {\Z}^{d}$ is governed by the
second-quantized Hamiltonian 
\beq
{\cal H}=- \sum_{x,y   \in \Lambda} t_{x,y} \,
c_{x}^{\dagger}c_{y} + 2\, U \sum_{x \in \Lambda} w_x \,
c_{x}^{\dagger}c_{x} \label{H1}
\eeq
where $c^{\dagger}_{x}$ and $c_{x}$ are the creation and annihilation
operators for the electrons at the site {\it x}, and $t_{x,y}$ is the
matrix element for hopping between two sites. 
Each lattice site may be occupied by at most one fixed nucleus which 
interacts with the mobile electrons via a on-site interaction {\it U}. The
interaction is repulsive when $U >0$ and attractive if $U <0$. The occupation
number $w_x$ is $0$ or $1$ according to whether the site {\it x} is
occupied or not by a nucleus. 

Although the $w$ variables are referred as nuclei, this
description is 
only one of many different model's interpretations. One may think of
the  
Hamiltonian (\ref{H1}) as an approximation to the
Hubbard model in which  one of the two types of electrons are kept static. See
references \cite{KL,K} for a survey on this.     

Compared to the Hubbard model, the Falicov--Kimball model has
the advantage to be reducible to a single particle Hamiltonian. Since there
are no interactions between the electrons, (\ref{H1}) is the second quantized
form of the Schroedinger Hamiltonian on $\ell _2 (\Lambda)$:
\beq
\label{H2}
H = - T + 2 \, U \; W
\eeq
where $T$ is the self--adjoint operator with matrix elements $t_{x,y}=t_{y,x}$
and $W$ is the multiplication operator by $w_x$, i.e., the diagonal matrix $W
= {\rm  diag}\left(\{w_x \}_{x\in \Lambda} \right)$. Denoting $L$ the number of
sites in $\Lambda$ ($L = |\Lambda |$), $T$ and $W$ are $L\times L$ matrices.  

To discuss some of the known results on the ground state of
(\ref{H1}), let $\Lambda$ be the union of two disjoint sub-lattices $\calA$
and $\calB$ 
and let us assume that $t_{x,y}$ vanishes when $x$ and $y$ belongs to the same
sub-lattice. It is also required  that $\Lambda$ is $T$--connected in the sense
that  $T$ is an irreducible matrix (it cannot be written as a direct sum). 
For example, these conditions are met if $T$ is the usual nearest neighbor
matrix  
\beq
\label{nn}
t_{x,y} = \left\{
\ba{lll}
t &  &{\rm if} \; |x-y|=1 \\
0 &  &{\rm otherwise}
\ea
\right. 
\eeq
$t>0$, since any $\Lambda \subset \Z ^d$ connected by their links is 
$T$--connected. Note that $T$ defines a bipartition of $\Lambda $: $x= (x_1,
\cdots , x_d)$ belongs to $\calA$ or $\calB$ according to whether  $x_1+
\cdots +x_d$ is an even or odd number.   

Now we introduce some notations e definitions. Let
\beq
N_{e}=\sum_{x \in \Lambda}c_{x}^{\dagger}c_{x} \;\;\;\;  {\rm and} 
\;\;\;\;  N_{n}=\sum_{x \in \Lambda} w_x
\eeq
denote the total number of electrons and the total number of nuclei,
respectively, and set  $\calN = N_e + N_n$. For each fixed configuration of
nuclei  $w= \{w_x \}_{x\in \Lambda}$ and  
$N_e$, $\calH$ has a ground state energy $E  (w,N_e)$. From this we define
two other kinds of ground state energies:
\beq
\label{gs}
E ^{(2)}(N_n,N_e) = \min _{w:N_n \;{\rm fixed}} \, E (w,N_e)
\eeq
and
\beq
\label{gs1}
E ^{(1)}(\calN) = \min _{N_n, N_e: \calN \;{\rm fixed} } \, E
^{(2)}(N_n,N_e)\; . 
\eeq
We shall also denote by $w_Q, \, Q\subset \Lambda$ the configuration of nuclei
with $w_x = 1$ if $x\in Q$ and $0$ otherwise.

Kennedy and Lieb \cite{KL} have proven the following
inequality
\begin{equation}\label{E12}
E ^{(1)} \ge  - \frac 12 \; {\rm Tr} \left( T^2
+ U^2 \right)^{1/2} + U\, \calN - \frac 12 \, U \, |\Lambda | 
\end{equation}
and concluded, from this, the following statements about the ground state of
(\ref{H1}):

\begin{description}
\item[A.]  Let $(U, \calN)$ be such that $U >0$ and  $\calN \ge |\Lambda |$, 
with $|\Lambda |$ the number of sites in $\Lambda $. Then the ground state
energy $E ^{(1)}$ has a minimum value which saturates the inequality 
(\ref{E12})  
at exactly the boundary $\calN = |\Lambda |$. The ground state is doubly
degenerated with the minimum of $E $  attained at either 
$(w, N_e)= (w_{\calA}, |\calB|)$ or $(w, N_e)= (w_{\calB}, |\calA|)$.

\item[B.] Let  $(U, \calN)$ be such that $U <0$ and  $\calN \le 2\, |\calA
|$. Then the ground state energy $E ^{(1)}$ has a minimum value given (\ref{E12}) 
at exactly the boundary $\calN = 2 \,|\calA|$. The minimum of $E
^{(2)}$ is attained at  $N_{n} = N_{e}= |\calA|$. The ground state is
unique unless $|\calA|=|\calB| = |\Lambda |/2$.
\end{description}

\begin{remark}
Notice that, if $|\calA|\not =|\calB|$, the ground state energy has an
asymmetric behavior when it passes from the {\bf A} to the {\bf B} condition
These two ground state energies coincide with 
(\ref{E12}) at $\calN 
=|\Lambda |$ only when $|\calA|=|\calB|=|\Lambda |/2$. In this 
situation the minimum of $E ^{(2)}$ is attained at $N_n =N_{e}= |\Lambda
|/2$. The ground state is doubly degenerated at the nuclei configuration
$w_{\calA}$ and $w_{\calB}$. The point $N_{e}= |\Lambda |/2 $ is called {\sl
half--filled band} 
because half of the available eigenstates of $H$ are filled with electrons.
Due to the fact that nucleus and electron 
have opposite charges, $N_{n}=N_{e}$ is often said to be the {\sl neutral
point} of $E ^{(2)}$. It is  also worth of attention that, when $T$
is given by (\ref{nn}), the nuclei occupy alternate sites on
the lattice forming a checker-board. We stress that {\bf A} and {\bf B}
results are independent 
of the hopping matrix entries (irrespective to their signs and strengths)
provided $t_{x,y}$ vanishes if $x$ and $y$ belongs to 
the same sub-lattice.   
\end{remark}

Kennedy and Lieb \cite{KL} have also proven existence of energy gap at
half--filled band. They
propose two gap definitions according to whether (\ref{gs1}) or
(\ref{gs}) is chosen as the ground state energy. Let $ \mu (\calN ) := E
^{(1)} (\calN +1) - 
E ^{(1)} (\calN )$ be the  chemical potential and let 
\beq
\label{gap1}
d ^{(1)} (\calN ):= \mu (\calN ) - \mu (\calN -1) = E ^{(1)}(\calN +1)
+ E ^{(1)} (\calN -1) - 2 E ^{(1)} (\calN ) 
\eeq
be the chemical potential discontinuity at $\calN$.  
The system is said to have a {\sl gap of the first kind} at
$\calN$ 
if there exist a constant $\varepsilon _1 >0$, uniformly in $\Lambda $, such
that $d ^{(1)} \ge \varepsilon _1$. Analogously, the gap is said to be of
{\sl 
second kind} at $(N_n, N_e)$ if 
\beq
\label{gap2}
d ^{(2)} (N_n, N_e) := E ^{(2)} (N_n, N_e +1) + E ^{(2)}(N_n, N_e
-1) - 
2 E ^{(2)} (N_n , N_e ) \ge \varepsilon _2 
\eeq
for some constant $\varepsilon _2 >0$, uniformly in $\Lambda $.

\begin{description}
\item[C.]  Let $T$ be an irreducible matrix such that $\| T \| \le\tau$ and
$t_{x,y}\ge \delta $ for all non--vanishing entries, uniformly in $\Lambda $,
for some constants $\tau <\infty $ and $\delta >0$. Under the
conditions of item {\bf A}, there is a first kind gap at $\calN = |\Lambda |$
and  second kind gaps at  $(N_n ,N_e)= (|\calA|, |\calB|)$ and at $(N_n ,N_e)=
(|\calB|, |\calA|)$ with $\varepsilon _2 \ge \varepsilon _1 >0$ depending only
on $\delta, \, \tau, \, U$. 

\item[D.] Let $T$ be as before. Under the
conditions of item {\bf B}, there are first kind gaps at $\calN = 2 \,|\calA |$
and at $\calN = 2 \,|\calB |$
and a  second kind gaps at  $(N_n ,N_e)= (|\calA|, |\calA|)$ and at $(N_n
,N_e)= (|\calB|, |\calB|)$ with $\varepsilon _2 \ge \varepsilon _1 >0$
depending only on $\delta, \, \tau, \, U$.
\end{description}

The Falicov--Kimball model remains object of intense investigation. For a
comprehensive survey of this model we quote a recent review by Gruber and
Macris \cite{GM} and references therein. Up to the present time, very few
results about the ground state do 
not require half--filled band or neutral point condition. Other model
restrictions which are frequently assumed
concern with the spatial dimension $d$ and the strength of interaction
$U$. The model which has been investigated in more detail has $T$ given by
(\ref{nn}) with $\Lambda \subset \Z$ ($d=1$) (see \cite{GM} for many
intriguing questions on this).  Also sufficiently large $|U|$ are assumed
very often. There are not many results when $T$ does not vanish at same
sub-lattice.  

In view of this scenario it seems 
worth to investigate one particular model where the eigenvalues and
eigenvectors of $H$ can be computed explicitly for all $\Lambda$. This program
can be accomplished if the hopping matrix $T$ has all non diagonal elements
equal to a constant. We call this hopping matrix {\sl mean field matrix} in
analogy to what is usually called mean field in classical spin systems. We
warn the reader that the Falicov--Kimball model with mean field hopping matrix
cannot be confused with the mean field approximation of the Hubbard model. 

The model Hamiltonian (\ref{H1}) with mean field hopping was introduced and
solved by Farkas\u ovsky \cite{F}, who called it simplified Hubbard model with
unconstrained hopping of electrons. Its
Hamiltonian is given by (\ref{H1}) or equivalently by (\ref{H2})
with the hopping matrix given by ($| \Lambda |=L$)
\beq
\label{lr}
t_{x,y} = \frac {2\,t}L
\eeq
for all $x,y \in \Lambda$ with $x\not = y$ and $0$ otherwise. 

Farkas\u ovisky's solution of this model is obtained 
by a canonical transformation $\calU$ that diagonalizes $H$. Unfortunately,
there is no mention in reference \cite{F} on how could $\calU$ be
computed. So, one of the purposes of this paper is to present a simple
derivation of the eigenvalue problem of $H$. 

What makes (\ref{lr}) solvable is the fact that the energy does not depend on
the nuclei configuration $w=\{w_x \}_{x\in \Lambda}$:  
\beq
\label{wind}
E (w, N_e) = E ^{(2)} (N_n, N_e)
\eeq
since the probability of hopping from any site $x$ to any other site $y$ is
evenly distributed. In view of this fact the model is not suitable to describe
the positional order of nuclei induced by 
electrons (cristalization). However many questions about the ground state
energy $E ^{(1)}$, as a function of $\calN$, can be formulated and
answered. We believe that these answers expresses general features of the
Falicov--Kimball model. We also believe that the solution of the eigenvalue
problem can be useful in studying disordered systems.   

Contrary to reference \cite{F} we have
made the hopping matrix size dependent ($t_{x,y}\rightarrow 0$ as $L\to
\infty$) in order to guarantee the
existence of thermodynamic limit. This simple fact 
allow us a complete description of the ground state for any value of the
model parameters  $U$, $L$ and $\calN$ (Farkas\u ovsky discuss only the
repulsive $U>0$ case for finite $L$).

The following summarizes our analysis.

\begin{theorem}
\label{gg}
Let $\calH$ be given by (\ref{H1}) with the hopping matrix $T$ given by
(\ref{lr}). Then, for $L > L_0$ with $L_0$ sufficiently large
depending on $U$, we have:
\begin{description}
\item[I.] If $U >0$, the ground state energy $E ^{(1)}$ is a monotone
increasing function of $\calN$ for all $0 < \calN \le 2L$. The minimum value of
$E ^{(2)}$ is attained at the following 
points:
\begin{itemize}
\item[a.] $(N_n , N_e ) = (\calN -1, 1)$ if $0 < \calN \le \rho ^\ast L$ with
$\rho ^\ast = \min \left(1/2 + 1/{(4u)} \, , \, 1 \right)$;

\item[b.] $(N_n , N_e ) = ([\rho ^\ast L], \calN -[\rho ^\ast L])$ if $[\rho
^\ast L] \le \calN \le L$ ($\, [x]$ means the integer part of the number $x\in
\R \,$); 

\item[c.] $(N_n , N_e ) = (L, \calN -L)$ if $L < \calN \le 2L$;

\end{itemize}

Moreover, $E ^1$ is constant in the domain $[\rho ^\ast L] \le \calN \le L$
(case b).

\item[II.] If $U<0$, the ground state energy $E ^{(1)}$ is a monotone
decreasing function of $\calN$ for all  $0 < \calN \le 2L$. Except by $1/L$
corrections, the minimum value
of $E ^{(2)}$ is attained at the neutral point $(N_n , N_e ) = (\calN
/2 , \calN /2 )$ 
\end{description}
\end{theorem} 

\begin{remark}
Monotone behavior of the ground state energy  $E ^{(1)}$ seems, from our
analysis, typical of the Falicov--Kimball model. This property would explain
the minima tendency to be attained at the domain boundary in {\bf A} and {\bf
B} results. The derivation of this property for nearest neighbor hopping
matrix in $d=1$ dimension is subject of a forthcoming paper \cite{MB}.
\end{remark}

\begin{theorem}\label{GAP}
Under the condition of Theorem \ref{gg}, the mean--field Falicov--Kimball
model has a first kind gap $d^{(1)}=2U$ at the half--filled band $\calN = L$,
if $U>0$ but no gap of first kind occurs if $U<0$.

In addition, there are three gaps of second kind at $(N_{n}, 1),(N_{n}, L-
N_{n}) $ and $(N_{n}, L- N_{n}+1)$ for all nuclei number $0\le N_{n} \le L$ if
$U>0$ and at $(N_{n}, 1),(N_{n}, N_{n}) $ and $(N_{n}, N_{n}+1)$, $0\le N_{n}
\le L$,  if $U<0$. The gap values are listed in (\ref{d>}) and (\ref{d<}).
\end{theorem}

\begin{remark}\label{nongap}
Absence of energy gap when $U<0$ does not contradict Kennedy--Lieb's
result. We note that the elements of the mean--field hopping matrix are not
uniform in 
$\Lambda $. We also observe that this hopping matrix violates the bipartition
condition required in item {\bf C}. As we shall see, this condition does not
affect our results on the gap which clearly indicate the existence of a phase
transition.  
\end{remark}

Finally, we illustrate the Kennedy--Lieb results state above with an
exactly solvable model. Our lattice $\Lambda$ may be arbitrarily 
decomposed 
into two sub-lattices  $\calA$ and $\calB$. The model, called by us bipartite
mean field, has non--vanishing hopping matrix elements $t_{x,y}$ equal to a
constant if $x,y$ 
connects the two sub--lattices. The Hamiltonian is thus given by (\ref{H1})
with 
\beq
\label{lr1}
t_{x,y} = \left\{
\ba{lll}
 \frac {2\,t}L & {\rm if} & x, y \; {\rm is \; not \; in \; the \; same \;
sub-lattice} \\
0 & & {\rm otherwise}
\ea
\right.
\eeq
The results of itens {\bf A} -- {\bf D} may be appreciated in light of
the eigenvalue problem for the present model. In particular, one sees that the
equation (\ref{E12}) becomes equality if one of the the sub--lattices, let
say $\calA $,  is constrained to be 
filled with nuclei ($|\calA |=N_{n}$) for $0 \le \calN \le 2L$ when $U<0$ and
for  $|\calN /L -1|  \ge  1/2$ when $U>0$.

This paper is organized as follows. In Section \ref{EP} the
eigenvalue problem of $H$ with mean--field hopping matrix (\ref{lr}) is
solved. Sections 
\ref{groundstates} and \ref{gap} give a complete description of the ground
state energy and energy gaps for this particular model. This analysis is
repeated for the bipartite mean--field model in Section \ref{bmfm}. We finally
draw some conclusions in the Section \ref{conclusions}.  Appendix \ref{prop1}
reviews some basic features about circulant matrices used in this text.


\section{The Eigenvalue Problem}\label{EP}
\zerarcounters

This section is devoted to the eigenvalue problem of the Schroedinger operator
(\ref{H2}) with mean field hopping matrix (\ref{lr}). 

For convenience, we write the interaction constant $U= t\; u$ and factorize
$2\, t$ from the Hamiltonian matrix $H$. This factor is just a scale to the
energy and may be fixed equal to $1$ without any loss of generality. Equation
(\ref{H2}) reads
$$
H= - T + u \, W \; .
$$
$H=H(w)$ is a $L\times L$ matrix with matrix elements $h_{x,y} = -1/L$
if $x\not = y$ and $h_{x,x} = u \, w_x$ which depends on the nuclei
configuration $w$. 

Now we claim that the spectrum of $H$ given by the zeros of the
characteristic polynomial
\beq
\label{cf}
P(\lambda )= \det \left(H - \lambda \, \I \right)
\eeq
is independent of the nuclei configuration. 

To see this, let $w$ and $w^\prime$ be two nucleus configurations differing by
a permutation $\pi$, i.e., $w^\prime _x= w_{\pi (x)}$. If $\Pi$ denotes the
permutation matrix determined by $\pi$, we have $\Pi ^{-1} \, T \, \Pi= T$ and 
\beq
\label{P}
H(w^\prime ) = \Pi ^{-1} \, H(w) \, \Pi \, .
\eeq
Note that the nearest neighbor matrix $T$ given by (\ref{nn}) is not invariant
by the similarity transformation (\ref{P}).  

As a consequence of this fact $\det \left(H(w^\prime ) - \lambda \, \I 
\right)= \det \left(H(w) - \lambda \, \I \right)$ and the
eigenvalues of $H$ depend only on the total number of nuclei $N_n$. We shall
drop the subscript $n$ of $N_n$ in the sequel of this section.

To compute (\ref{cf}) let us pick the simplest matrix $H$. For any positive
integers $L$ and $N$, $L > N$,  let $J_1, J_2 $ and $K$ be 
given by the following matrices
\beq
\label{matrices}
\ba{lll}
J_1 & = & \left(u + 1/L \right) \, I_N - (1/L) \, \UM _N\\
J_2 & = &  (1/L) \, I_{L-N} - (1/L) \, \UM _{L-N}\\
K & = & - (1/L) \, \UM _{N,L-N} 
\ea
\eeq
where $I _R$ is the identity matrix of size $R$, $\UM _{R,S}$ is the $R\times
S$ matrix with all elements equal to $1$ and  $\UM _{R} =\UM _{R,R} $. We set
\beq
\label{H0}
H_0 =
\left (  
\begin{array}{cc}
J_1 & K \\
K^{T}& J_2 \\
\end{array}
\right ) \; .
\eeq
Note that $H_0 = H(w_0)$ with the nuclei configuration $w_0= (1, 1, \dots, 1,
0, \dots , 0)$.
 
In view of the above,  the spectrum of $H(w)$ can be read from the spectrum of
$H_0$ given as follows:

\begin{proposition} \label{eigenvalues}
The spectrum of $H_0$ consists of the set $\{ \lambda _1, \lambda _2, \lambda
_+, \lambda _-\}$ where
\beq
\label{sp}
\ba{lll}
\lambda _1 & = & 1/L \\
\lambda _2 & = &  1/L + u \\
\lambda _\pm & = & 1/L +  \left[u-1 \pm \Delta (N/L)\right]/2 
\ea
\eeq
with
\beq
\Delta (\rho) :=  \sqrt{(u+1)^{2}- 4 u \rho} \; ,
\eeq
have multiplicities $L-N-1$, $N-1$ and $1$, respectively. 

The corresponding
eigenvectors ${\bf x}= ({\bf u}, {\bf v }) \in \C ^N \times \C ^{L-N}$ are 
${\bf x}_{1,j}=({\bf u}_{N,j}\, , \,{\bf 0})\; , j=1, \dots , N-1$, ${\bf x}_{2,k}=({\bf 0} \, ,
\,{\bf u}_{L-N,k})\; , k=1, \dots , L-N-1$, and ${\bf x}_\pm =({\bf 1} \, , \,
{\bf u}_{\pm})$ 
where ${\bf 0}= (0, \dots  , 0)$ is the zero--vector,  ${\bf 1}= (1, \dots  ,
1)$ is the one--vector,
\beq
\label{u+-}
{\bf u}_{\pm}=\frac{u+1-2 N/L \pm \Delta (N/L)}{2(1-N/L)} \, {\bf 1}
\eeq
and ${\bf u}_{R,j}=(1, \omega_R^{j}, \omega_R^{2j}, \dots  ,
\omega^{N j}_{R}), j =1, \dots R-1$, with $\omega _R =e^{2 \pi
i/R}$. 
\end{proposition}

The proof of Proposition \ref{eigenvalues} will follow from two lemmas. The
first reduces the problem to matrices of the form $\alpha \, I+ \beta
\,\UM$. The second describes the properties of such matrices.

\begin{lemma}
\label{red}
\beq
\label{red1}
\det \left(H_0 -\lambda I \right) = \det \left(J_1 -\lambda I \right) \cdot
\det \left[\left(J_2 -\lambda I
\right) - K^T \left(J_1 -\lambda I \right)^{-1} K\right] \, .
\eeq
\end{lemma}

$\Proof$  Let
$$
M= \left(  
\begin{array}{cc}
A & K \\
K^{T}& B \\
\end{array}
\right),
$$
be such that  $A$, $B$ and $K$ are $N\times N$, $(L-N)\times (L-N)$ and $N
\times (L-N)$ matrices, respectively, with $A$ symmetric and invertible. $M$
can be brought to a block diagonal form  by using a non--unitary
transformation  
\beq
\label{block}
\left (
\begin{array}{cc}
A & 0 \\
0 & B-K^T A^{-1}K \\
\end{array}
\right ) =
\left (
\begin{array}{cc}
1 & 0 \\
-K^{T}A^{-1} & 1 \\
\end{array}
\right )
\left(  
\begin{array}{cc}
A & K \\
K^{T}& B \\
\end{array}
\right)
\left (
\begin{array}{cc}
1 & - A^{-1}K \\
0 & 1 \\
\end{array}
\right ) \, .
\eeq

Taking the determinant of both sides of equation (\ref{block}) with $A= J_1
-\lambda I$ and $B=J_2 -\lambda I$ gives (\ref{red}). 
$\QED$

\begin{lemma} \label{prop} \begin{enumerate}
\item The linear space $\calC$ of all $R\times R$ matrices of the form $\alpha
\, I+ \beta \,\UM$ is closed under sum, product and inverse operations. 

\item The inverse of a matrix $C =\alpha \, I+ \beta \,\UM$ is
given by
\beq
C^{-1} = \frac 1\alpha \, I - \frac \beta{\alpha(\alpha + \beta R)} \, \UM
\eeq
provided $\alpha \not =0$ and $\alpha \not = -\beta R$.

\item  Any  matrix $C =\alpha \, I+ \beta \,\UM$ can be reduced to a diagonal
form 
$$
F^{-1} \, C \, F = {\rm diag}\left\{\alpha + \beta R, \alpha, \dots, \alpha
\right\} 
$$
where $F$ is the Fourier matrix 
$$
F_{k,l} := \frac 1{\sqrt R} \, \omega _R^{(k-1)(l-1)}  \; , \;\;\;\;\;\;\; k,l =
1, \dots R 
$$
with $\omega _R= e^{2\pi\, i/R}$.

In other words, $C$ has a simple eigenvalue $\lambda _1 = \alpha + \beta N$ 
corresponding to the eigenvector ${\bf 1}= (1, 1, \dots, 1)$ and a
$(R-1)$--fold 
eigenvalue $\lambda _2 = \alpha$ associated to eigenvectors ${\bf u}_{R,j} =
(1, \omega _R^j , \omega _R^{2j}, \dots, \omega _R^{(R-1)j})$, $j=1, \dots ,
(R-1)$. 

\item $\det C = (\alpha + \beta R)\, \alpha ^{R-1}$.
\end{enumerate}
\end{lemma}

The proof of Lemma \ref{prop} is in the Appendix \ref{prop1}.

\noindent 
{\it Proof of Proposition \ref{eigenvalues}} We use Lemma \ref{red} to reduce
the calculation of the characteristic function $P_0(\lambda )= \det (H_0
-\lambda I)$ to algebraic calculations of matrices of $\calC$. 

Using item {\it 1.} and {\it 2.} of Lemma \ref{prop}  and definitions
(\ref{matrices}), we have 
$$
\left(J_1 -\lambda I \right)^{-1} = \frac 1{u + 1/L -\lambda} \, I_N + \frac
{1/L}{(u + 1/L -\lambda )(u -(N-1)/L -\lambda )} \, \UM _N
$$
and, in view of $\UM _{R,S}\, \UM _{S,R}  = S \, \UM _R$,
$$
K^T \left(J_1 -\lambda I \right)^{-1} K =  \dfrac {N/L^2}{u -(N-1)/L
-\lambda}\, \UM _{L-N} 
$$
which gives
\beq
\label{det}
(J_2 - \lambda I) - K^T \left(J_1 -\lambda I \right)^{-1} K = (1/L -\lambda )
\, I_{L-N} - \frac {(u + 1/L -\lambda )/L}{u - (N-1)/L -\lambda }\, \UM _{L-N}
\, .
\eeq

We then use item {\it 4.} of Lemma \ref{prop} and equations (\ref{matrices})
and (\ref{det}) to write
\beq
\label{det1}
\det (J_1 -\lambda I) = \left( u -(N-1)/L -\lambda \right)\, \left( u + 1/L
-\lambda \right)^{N-1}
\eeq
and, with a little of algebra,
\beq
\label{det2}
\det \left( (J_2 - \lambda I) - K^T \left(J_1 -\lambda I \right)^{-1} K
\right)= \frac {(1/L -\lambda )^{L-N-1}}{u - (N-1)/L -\lambda } \, R(\lambda)
\eeq
where
$$
\ba{lll}
R(\lambda ) & = & \lambda  ^2 - (u-1-2/L)\lambda -u(L-N-1)/L -1/L +1/L^2 \\
 & = & (\lambda - \lambda _+)(\lambda - \lambda _-) \, .
\ea
$$

The eigenvalues of $H_0$ then follows from (\ref{det1}), (\ref{det2}) and
Lemma \ref{red}.

The corresponding eigenvectors of $H_0$ can be obtained from item {\it 3.} of
Lemma \ref{prop}. A vector ${\bf x}$ in $\C ^L$ is written as $({\bf 
u},{\bf v})$ where ${\bf u}\in  \C ^N$ and ${\bf v}\in  \C ^{L-N}$.  Using
Lemma \ref{prop} and the notation of Proposition \ref{eigenvalues}, the
eigenvectors corresponding to the $\lambda _1$ and $\lambda _2$ are, by
inspection,  
given by $({\bf u}_{N,j},{\bf 0}), j=1, \dots , N-1$ and $({\bf 0},{\bf
u}_{L-N,k}), k=1, \dots , L-N-1$, respectively. Notice $K\, {\bf u}_{N,j} =
{\bf 0}$ for all $j=1, \dots , N-1$ in view of the property (\ref{0}) in the
Appendix \ref{prop1}. By inspection also, one can
verify that the eigenvectors corresponding to $\lambda _\pm$ are $({\bf 1},
{\bf u}_\pm)$ with ${\bf u}_\pm$ given by (\ref{u+-}). 

This concludes the proof of Proposition \ref{eigenvalues}.
$\QED$



\section{Ground States}
\label{groundstates}
\zerarcounters

For fixed nuclei configuration $w$, let $\mu _1 \le \mu _2 \le \cdots \le \mu
_L$ be the sequence of ordered 
eigenvalues of (\ref{H2}), counting multiplicities, and let $N_e$ be the total
number of electrons in the system.  

Due to the exclusion principle, each energy level (eigenvalues $\mu _j$) can
be occupied by at most one electron. In the ground state the first $N_e$ lowest
energy levels of $H$ are occupied by electrons and the $L - N_e$
remaining levels are left empty. The ground state energy $E$ of the model
Hamiltonian (\ref{H1}) at $(w,  N_e)$ is thus determined by the sum of the
lowest eigenvalues: 
\beq
\label{E}
E (w, N_{e}) := \sum _{j=1}^{N_e}
\mu _{j} \, . 
\eeq

We shall here prove Theorem \ref{gg} on the ground state energy $E ^{(1)}$
of the Hamiltonian (\ref{H1}) with mean field 
hopping matrix (\ref{lr}) whose eigenvalues have been studied in the previous
section.  

Since we have proven the spectrum of $H$, $\sigma (H) = \{\lambda _1 , \lambda
_2 ,\lambda _+ ,\lambda _- \}$, depends only on the total number of nuclei, we
have   
$$
E ^{(2)}(N_n, N_{e}) = E (w, N_{e})
$$
for all $w$. Consequently, we shall only be concerned with the minimum value
of  $E ^{(2)}$ for fixed number of particles $\calN = N_n + N_e$. 

Our discussion about ground state energy begins by the eigenvalues
ordering. We observe that the spectrum $\sigma (H)$ depends on the nuclei
density 
$\rho _n =  N_n /L$ through the function  
\beq
\label{Delta}
\Delta ^2 (\rho ) := (u+1)^2 -4 \, u \, \rho 
\eeq
which, in view of $0 \le \rho  \le 1$, satisfies
$$
\left(|u| -1\right)^2 \le \Delta ^2 \le \left(|u| + 1\right)^2 \, .
$$

Since $\lambda _\pm = 1/L + (u-1 \pm \Delta )/2$, this implies 
$$
1/L +  \left( u-1 + \left|\, |u| - 1 \, \right| \right)/2 \le
\lambda _+ \le 1/L + 
\left(u + |u|  \right)/2 
$$
and 
$$
\lambda _- \le 1/L + \left( u-1 - \left| \, |u| - 1 \,\right| \right)/2 
$$
from which the eigenvalues ordering follows. 

We distinguish two cases: 
\begin{description}
\item[I.] if $u> 0$ we have 
$$
\lambda _- \le \lambda _1 \le \lambda _+ \le \lambda _2 \, ;
$$

\item[II.] if $u< 0$ we have
$$
\lambda _- \le \lambda _2 \le \lambda _+ \le \lambda _1 \, .
$$
\end{description}

\noindent
{\it Proof of part {\bf I} of Theorem \ref{gg}.} Let $u>0$ and $\calN \le
L$. In this case we have 
\beq
\label{mumu}
\mu _1 =\lambda_- \;  , \;\; \mu _2 = \cdots  =\mu _{L-N_n} = \lambda_1 \; ,
\;\; \mu _{L-N_n +1} =\lambda_+ \;  , \;\; \mu _{L-N_n +2} = \cdots  =\mu _{L}
= \lambda_2 
\eeq
and the ground state energy is thus given by
\beq
\label{GS1}
\ba{lll}
E ^{(2)} (N_n, N_e) & = & \lambda _- + (N_e -1) \, \lambda _1 \\
 & = & \lambda _- + (\calN - N_n -1) \, \lambda _1 \\
 & = & \calN /L + g(\rho _n)
\ea
\eeq
where
\beq
\label{g}
g(\rho ) := \frac 12 \left( u-1 - \Delta (\rho ) \right) -
\rho \,  
\eeq
and $\Delta $ as in (\ref{Delta}). We remind that, under the 
condition $\calN \le L$, the multiplicity $L- N_n -1$ of $\lambda _1$ is
greater than or equal to the number of 
electrons $N_e -1$. This explain why we have taken only two eigenvalues in
(\ref{GS1}). 

Equation (\ref{GS1}) and (\ref{g}) reduces the problem of minimizing  $E
^{(2)}(N_n, N_e)= E ^{(2)}
(N_n, \calN  -N_n)$ with respect to $N_n$ to the problem of minimizing
$g$ as a function of $\rho _n$. The function $g : [0,1] \rightarrow \R$  has
the following properties: 

\begin{enumerate}

\item $g(\rho ) \le g(0)= -1$ with $g(1) = -1$ if $u\ge 1$;

\item $g$ attains to a unique minimum value at  
$$
\rho ^\ast = \min \left(\frac 12 + \frac 1{4\,u}, 1 \right) \, ;
$$
(note that $\rho ^\ast = 1$ if $u \le 1/2$);

\item the minimum value is given by 
\beq
\label{rho*}
g(\rho ^\ast) = \left\{ 
\ba{lll}
u- 2 & {\rm if} & 0 < u < 1/2 \\
-1 - 1/(4\, u) & {\rm  if}  & u \ge 1/2 \, ,
\ea
\right.
\eeq

\end{enumerate}

Items $1 , 2$ and $3$  can be obtained explicitly. Note that the derivative
\beq
\label{gp}
g^\prime (\rho) = \frac u{\Delta ( \rho )} -1 
\eeq
vanishes only at $\rho^\ast $ provided $u\ge 1/2$. If $u < 1/2$,  $g$ attains
its minimum value at the domain boundary $\rho =1$. 

From equation (\ref{GS1}) and properties $1 , 2$ and $3$ we conclude that the
ground 
state $E ^{(1)}(\calN)$ has only one eigenvalue $\lambda _-$ occupied
by one electron if $\calN \le \rho^\ast \, L$ and there will be more electrons
contributing to the ground state if  $\calN > \rho^\ast \, L$. Note that
$\lambda _- =\lambda _- (\rho _n)$ is  negative if $\rho _n \le  \rho ^\ast$
provided $L >2$. This follows from the fact that $\lambda _-
(\rho_{n}) \le \lambda _-
(\rho ^\ast) = 1/L + (u -1 - \Delta (\rho ^\ast))/2 = 1/L -1/2$ since, from
(\ref{gp}), $\Delta (\rho ^\ast) = u$. We shall always assume $L$ sufficiently
large.   

In summary, 
\beq
\label{GSIa}
E ^{(1)} (\calN) = 
\left \{
\begin{array}{lll}
\lambda _-\left((\calN -1)/L\right) & {\rm if} &
\calN < \rho ^\ast \, L  \\ 
\calN /L + g\left([\rho ^\ast\, L]/L \right) & {\rm if } &  \rho
^\ast \, L \le \calN \le L 
\end{array}
\right.
\eeq
where $[x]$ means the closest integer of  $x \in \R$. We 
emphasize that $E ^{(1)}(\calN)$ is a monotone increasing function of $\calN$,
since $\lambda _-$ is a monotone increasing function of the nuclei number, and
piecewise constant if $\rho ^\ast \, L \le \calN \le L$.

\medskip

Now, let $u>0$ and $\calN > L$. The ground state energy $E^{(2)}
(N_n, N_e)$ in this case is given by
\beq
\label{GS2}
\ba{lll}
E^{(2)} (N_n, N_e) & = & \lambda _- + (L -N_n -1) \, \lambda _1 + \lambda _+ +
(N_e - (L - N_n +1)) \, \lambda _2 \\
 & = & \lambda _- + (L -N_n -1) \, \lambda _1 + \lambda _+ +
(\calN -L -1) \, \lambda _2 \\
 & = & (\calN - N_n) /L + (\calN -L) u -1
\ea
\eeq
which attains to a minimum value at $(N_n, N_e)= (L, \calN -L)$ corresponding
to the maximal nuclear occupation. We thus have
\beq
\label{GSIb}
E^{(1)}  (\calN) = (\calN - L) (1/L + u) -1
\eeq
for $L < \calN \le 2L$. The combination of (\ref{GSIa}) and (\ref{GSIb}) leads
to a monotonic increasing ground state energy $E^{(1)}$ as a function of
$\calN$  in the whole domain.

This concludes the analysis of the case {\bf I}. $\QED$

\medskip

\noindent
{\it Proof of part {\bf II} of Theorem \ref{gg}.} Here, we have
\beq
\label{mumu1}
\mu _1 =\lambda_- \;  , \;\; \mu _2 = \cdots  =\mu _{N_n} = \lambda_2 \; ,
\;\; \mu _{N_n +1} =\lambda_+ \;  , \;\; \mu _{N_n +2} = \cdots  =\mu _{L}
= \lambda_2 \, .
\eeq

We shall divide our analysis in two sub-cases. Let $u < 0$ and $N_e\le
N_n$. In view of the  
eigenvalue ordering, 
we have
\beq
\label{GS3}
\ba{lll}
E^{(2)}  (N_n, N_e) & = & \lambda _- + (N_e -1) \, \lambda _2 \\
 & = & \lambda _- + (\calN - N_n -1) \, \lambda _2 \\
 & = & \calN /L + u (\calN -1) + h(\rho _n)
\ea
\eeq
where
\beq
\label{h}
h(\rho ) = g(\rho) - u L \rho \, . 
\eeq

Once again, the problem of minimizing $E^{{(2)}} (N_n, N_e)$ under the
condition $N_e\le N_n$ is reduced to the
problem of minimizing $h$ in the domain  $1/L \le \rho \le 1$. Note that the
lower limit $1/L$ is due to the fact the equation (\ref{GS2}) is valid only if
$1 \le N_e \le \rho _n\, L $. The following description of $h$ gives the answer
to this problem. 

\begin{proposition}
\label{mono}
Given $u <0$, there is $L_0 = L_0(u)$ such that, for all $L > L_0$, $h: [1/L
, 1]\rightarrow \R$ given by (\ref{h}) is a strictly monotone increasing
function of $\rho $.
\end{proposition}

\noindent
{\it Proof.} Proposition \ref{mono} is implied if $h^\prime (\rho) >0$ for all
$\rho$ in $[1/L , 1]$. From (\ref{h}) and (\ref{gp}) this is implied by
\beq
\label{c1}
\frac u{\Delta (\rho )} > uL + 1 \, .
\eeq

To show that condition (\ref{c1}) holds for any value $u \in (-\infty , 0)$, we
must break this open interval into three pieces: $(-\infty ,-1 -1/L) \cup  (-1
-1/L ,-1 +1/L) \cup(-1+ 1/L ,-1/L)$. Note that we have to take $L$ large in
order get $u$ arbitrarily close to $0$. This leads to $L > - 1/u$.  

Let us assume $-1 +1/L < u < -1/L$. Using the inequality $\Delta (\rho ) >
u+1$, equation (\ref{c1}) can be replaced by a more restrictive condition
$$
\frac u{u+1} > uL + 1 \, .
$$
This is equivalent to  $ L \, u^2 + L\,  u + 1 = L \,(u +1 -1/L)(u+1/L) < 0$
which is always true for $u \in (-1 +1/L , -1/L)$. 

Suppose now  $u <-1 - 1/L$. Then we analogously use  $ \Delta (\rho ) >
-(u+1)$ to replace (\ref{c1}) by a more restrictive 
condition  
\beq
\label{c2}
L \, u^2 + (L+2)  \, u + 1  > 0 \, .
\eeq
Since the  roots $u= - 1/2 -1/L \pm [1/4 +1/L^2]^{1/2}$ of the quadratic
equation are close to $-1$ and $0$, (\ref{c2}) is always satisfied if $u < -
1/2 -1/L - [1/4 +1/L^2]^{1/2} < - 1 -1/L$.

It remains to verify condition (\ref{c1}) for $-1 -1 /L < u < -1 + 1/L$. In this
case we use $\rho > 1/L$ to replace (\ref{c1}) by 
\beq
\label{c3}
\frac{u}{\Delta (1/L)}= \frac u{\sqrt{ (u+1)^2 -4u /L }} > uL + 1 \, .
\eeq
Since $u = {\calO} (1)$ and $(u+1) = {\calO} (1/L)$, the left hand side of
this inequality is ${\calO} \left( \sqrt L \right)$ and the right hand side
${\calO} 
\left( L \right)$. Recall $u\le -1/L$ and both sides of (\ref{c3}) are
negative. Therefore, we can always find $L$ sufficiently large so that 
condition (\ref{c1}) is verified concluding the proof of the
Proposition. $\QED$

Proposition \ref{mono} implies that the ground state energy 
$E ^{{(2)}}$ under the condition $N_e \le N_n$ has a minimum value at 
$N_n = N_e = \calN /2$ given by
\beq
\label{GSIIa}
E^{(1)} _< (\calN) = \calN /L + u (\calN -1) + h(\calN /(2L))
\eeq  

Finally, let us consider $u<0$ and $N_e > N_n$. The ground state energy in
this case is given by
\beq
\label{GS4}
\ba{lll}
E^{(2)} (N_n, N_e) & = & \lambda _- + (N_n -1) \, \lambda _2 + \lambda _+ +
(N_e - N_n -1) \lambda _1\\
 & = & \calN /L -1 + (Lu - 1) N_n /L
\ea
\eeq

Since $E^{(2)} (N_n, \calN -N_n)$ is a monotone decreasing function of $\rho
_n = N_n /L$, it attains to a minimum value at $N_n =N_e -1 = \calN - N_n
-1$ given by
 \beq
\label{GSIIb}
E^{(1)} _> (\calN) =  \calN /L -1 + (Lu - 1) (\calN -1)/(2L) \, .
\eeq

The ground state for the case {\bf II} is thus given by 
\beq
\label{GSII}
\begin{array}{lll}
E^{(1)} (\calN) & = &\min \left(E^{(1)} _< (\calN)\,  , \, E^{(1)} _> (\calN)
\right) \\
& = & \dfrac u2 (\calN -1) + \dfrac {\calN}{2L} - \dfrac m2 
\end{array}
\eeq
where $m:= \max \left(2-1/L , 1+\Delta (\calN / (2L)) \right)$.

Note that $E^{(1)}$ is a monotone decreasing function of $\calN$ if
$L$ is sufficiently large. We point out that the minimum of 
$E^{(2)}$ is attained at approximately neutral (symmetric)  point 
$N_n + \calO (1/L) =  N_e + \calO (1/L) = \calN /2$ for all values of $u<0$
and $0 \le \calN \le 2L$.  

This concludes the proof of Theorem \ref{gg}. $\QED$

\section{Gaps}
\label{gap}
\zerarcounters

Let $\{ e_{L} (\rho ) \}_{L\ge 0}$ be the  sequence of  ground state energy
density functions defined by
\begin{equation}\label{eL}
e_{L} (\rho) = \frac{E^{(1)} (\rho \, L)}{L} \, 
\end{equation}
and let  $e (\rho ) = \lim _{L\to \infty} e_{L} (\rho )$ be its thermodynamic
limit. 


From our previous analysis, the ground state energy $E^{(1)}$ can be written in the
form (see equations (\ref{GSIa}), (\ref{GSIb}) and (\ref{GSII}))   
\begin{equation}\label{form}
E^{(1)} (\rho \, L)=  a_{0} (u, \rho , L) + a_{1} (u, \rho )  + a_{2} (u, \rho
) \, L 
\end{equation}
where, for fixed $u$, $\; a_{0}, a_{1}$ and $a_{2}$ 
are bounded functions of $\rho \in [0,2]$ with $a_{0}\to 0$ as
$L\to \infty$. 

Hence, given two positive integer numbers $L, M$, $L<M$,
\begin{equation}\label{dif}
\left|\dfrac{E^{(1)} (\rho \, L)}{L} - \dfrac{E^{(1)} (\rho \, M)}{M} \right|
\le  \left| \dfrac {a_{0} (L)}{L} - \dfrac {a_{0} (M)}{M}  \right| + a_{1}
 \left| \dfrac{1}{L}- \dfrac {1}{M} \right| 
 \le  C \dfrac {1}{L}
\end{equation}
for some constant $C=C(u) < \infty$, uniformly on $L$, and this shows
that $\{ e_{L} \}_{L\ge 0}$ forms a Cauchy sequence which converges to $e 
(\rho )$. 

It also follows from (\ref{GSIa}), (\ref{GSIb}) and (\ref{GSII}) that: 
\begin{description}
\item [a.] if $u>0$, we have 
\begin{equation}\label{e>}
e (\rho )=\left\{ 
\begin{array}{lll}
0 & {\rm if} & 0 \le \rho \le 1 \\
(\rho -1) u & {\rm if} & 1 < \rho \le 2  \, .
\end{array}\right.
\end{equation}
( Note that $a_{2}=0$ when $0 \le \rho \le 1$); 

\item [b.] if $u<0$, we have $e (\rho )=u \, \rho /2 $ for all $0 \le \rho \le 2$.

\end{description}

By a similar computation, one can show that the sequence of chemical
potentials $\{ \mu _{L} (\rho ) 
\}_{L\ge 0}$ given by 
\begin{equation}\label{muL}
\mu _{L} (\rho) := E^{(1)} (\rho \, L+ 1) - E^{(1)} (\rho \, L) =  \dfrac
{e_{L} (\rho +1/L) -   e_{L} (\rho )}{1/L} 
\end{equation}
converges to the derivative of the ground state energy
density $e^{\prime} (\rho )$ for all $\rho \in [0, 1)\cap (1, 2]$ provided
$u>0$. As a consequence, there exists a positive gap of first kind at $\rho =1$:
\begin{equation}\label{g1}
\lim _{\rho \searrow 1}\mu (\rho ) -\lim _{\rho \nearrow 1} \mu (\rho ) = u
\,. 
\end{equation}

For $u<0$ the sequence of chemical potentials $\{ \mu _{L} (\rho ) 
\}_{L\ge 0}$, converges to $e_{L}^{\prime} (\rho ) = u/2$
for all $\rho \in [0,2]$ and there is no first kind gap in this regime.

Now, let us consider the second kind gap. From the definitions (\ref{gap2}) and
(\ref{E}) we have 
\begin{equation}\label{gap2a}
d^{(2)} (N_{n}, N_{e}) = \mu _{N_{e}+1} - \mu _{N_{e}} 
\end{equation}
(For this, recall (\ref{wind})).

For $u>0$, it follows from the equation (\ref{mumu}) several cases:
\begin{equation}
\label{d>}
\begin{array}{llllll}
1. \;\;\;\;\; \;     & d^{(2)} (N_{n}, 1) & = & \dfrac {-u +1 +\Delta (\rho
_{n})}{2} & &  \;\;\;\;\; \;\\  
2.\;\;\;\;\; \; & d^{(2)} (N_{n}, N_{e}) & = & 0 & {\rm  if} & 1 < N_{e}< L -
N_{n}  \;\;\;\;\; \;  \;\;\;\;\; \;\\ 
3.\;\;\;\;\; \; & d^{(2)} (N_{n}, L - N_{n}) & = & \dfrac {u - 1 + \Delta (\rho
_{n})}{2} & & \;\;\;\;\; \; \\ 
4.\;\;\;\;\; \; & d^{(2)} (N_{n}, L - N_{n}+1) & = &\dfrac {u + 1 - \Delta
(\rho _{n})}{2}    &  & \;\;\;\;\; \; \\ 
5.\;\;\;\;\; \; &  d^{(2)} (N_{n}, N_{e}) &=& 0 & {\rm if} & N_{e} >L - N_{n}
+1 \;\;\;\;\; \;  \;\;\;\;\; \;
\end{array}
\end{equation}

We thus find three gaps for $u>0$: second kind gaps occur at $N_{e} = 1,
L-N_{n}$ and $L-N_{n}+1$ for each $0\le N_{n} \le L$.

For $u<0$ equation (\ref{mumu1}) gives the following:
\begin{equation}
\label{d<}
\begin{array}{llllll}
1. \;\;\;\;\; \;     & d^{(2)} (N_{n}, 1) & = & \dfrac {u +1 +\Delta (\rho
_{n})}{2} & &  \;\;\;\;\; \;\\  
2.\;\;\;\;\; \; & d^{(2)} (N_{n}, N_{e}) & = & 0 &\;\;\;\;\; {\rm  if} & 1 < N_{e}< 
N_{n}  \;\;\;\;\; \;  \;\;\;\;\; \;\;\;\;\;\; \; \;\;\; \\ 
3.\;\;\;\;\; \; & d^{(2)} (N_{n}, N_{n}) & = & \dfrac {-u - 1 + \Delta (\rho
_{n})}{2} & & \;\;\;\;\; \; \\ 
4.\;\;\;\;\; \; & d^{(2)} (N_{n}, N_{n}+1) & = &\dfrac {-u + 1 - \Delta
(\rho _{n})}{2}    &  & \;\;\;\;\; \; \\ 
5.\;\;\;\;\; \; &  d^{(2)} (N_{n}, N_{e}) &=& 0 &\;\;\;\;\;  {\rm if} & N_{e} > N_{n}
+1 \;\;\;\;\; \;  \;\;\;\;\; \;\;\;\;\;\; \;\;\;\;
\end{array}
\end{equation}

From these we find that, for $u<0$, second kind gaps occur at  $N_{e} = 1,
N_{n}$ and $N_{n}+1$ for each $0\le N_{n} \le L$ and this concludes the proof
of Theorem \ref{GAP}.
$\QED$







\section{Bipartite Mean Field Model}
\label{bmfm}
\zerarcounters

We shall briefly in this section analyze the
ground state of the Hamiltonian $\calH $ with bipartite mean--field hopping
matrix $T$ given by (\ref{lr1}). 

We begin by solving the eigenvalue problem. 
Let us decompose the vector space $\C ^{L}$, $L=| \Lambda |$, into
two subspace $\C ^{N}$ and $\C ^{M}$ corresponding to each sub-lattice $\calA $
and $\calB $ where $N= | \calA |$ and $M= | \calB |$. The nuclei occupation
number in a sub--lattice $\calX \subset \Lambda $ will be denoted by
$n_{\calX }$. We shall frequently write $n = n_{\calA }$ and $m = n_{\calB
}$. From these notations we have $L=N+M$ and $N_{n} = n +m $.  

The Hamiltonian $H=H
(w)$, acting on $\C ^{L}$, can thus be written as a block matrix 
\begin{equation}\label{Hmf}
H= \left( 
\begin{array}{cc}
J_{\calA } & K \\
K^{T} & J_{\calB }
\end{array}
\right)
\end{equation}
where $J_{\calX }= u \; {\rm diag}\{w_{x}, x\in \calX  \}$, $\calX =\calA
,\calB $ and $K=- (1/L) \,\UM _{N,M} $. 

As before, we can find a permutation matrix $\Pi $ such that $H_{0}=\Pi ^{-1}H
\Pi $ has the diagonal block matrices $J_{\calX
}$ of the form ${\rm  diag}\{u, \dots , u, 0, \dots , 0 \}$. Hence, in
view of Lemmas \ref{red} and \ref{prop}, the
characteristic polynomial $P (\lambda )$ of $H_{0}$ is given by 
\begin{equation}\label{chP}
P (\lambda ) = \det \left(H_{0}-\lambda I \right) = \det A \, \det C 
\end{equation}
where $A = \left( J_{\calA }-\lambda I  \right)$ and $C = \left( J_{\calB
}-\lambda I  \right)- K^{T}\, \left(J_{\calA }-\lambda I  \right)^{-1}\,K$,
can be written as 
\begin{equation}\label{C}
C = \left( 
\begin{array}{cc}
E & G \\
G^{T} & F
\end{array}
\right)
\end{equation}
with
\begin{equation}\label{EFG}
\begin{array}{lll}
E & = & (u- \lambda ) \, I_{m} - \alpha \, \UM _{m} \\
F & = & - \lambda  \, I_{M-m} - \alpha \, \UM _{M-m} \\
G & = &  - \alpha \, \UM _{m, M-m}
\end{array}
\end{equation}
and 
\begin{equation}\label{alpha}
\alpha = \frac{1}{L^{2}} \frac{(u-\lambda )\, N - u\, n}{\lambda \, (u-\lambda
)} \, .
\end{equation}

We now repeat the calculation given in the proof of Proposition
\ref{eigenvalues}. Once again, by Lemmas \ref{red} and \ref{prop} 
\begin{equation}\label{C1}
\det C = \det E \, \det H
\end{equation}
where 
\begin{equation}\label{H}
\begin{array}{lll}
H & = & F - G^{-1} E^{-1} G \\
 & = & -\lambda \, I_{M-m} - \dfrac{\alpha (u-\lambda )}{u-
\lambda -\alpha \, m} \, \UM _{M-m}
\end{array}
\end{equation}

Putting all pieces together, by using item {\it 4} of Lemma \ref{prop} we have
\begin{equation}\label{P2}
\begin{array}{lll}
\det A & = & (u-\lambda )^{n} \, (-\lambda )^{N-n} \\
\det E & = & (u-\lambda )^{m-1} \, (u-\lambda -\alpha \, m) \\
\det H & = & - (-\lambda )^{M-m-1} \; \dfrac{\lambda (u- \lambda )+ \alpha [u
(M-m)-\lambda \, M]}{u- \lambda - \alpha \, m} \, .
\end{array}
\end{equation}

The characteristic polynomial $P (\lambda )=
\det A  \, \det E \, \det H $ can thus be written as
\begin{equation}\label{P3}
P (\lambda ) = (u-\lambda )^{N_{n}-2}\, (-\lambda )^{L-N_{n}-2}\, S (\lambda )
\end{equation}
(recall $L= N+M$ and $N_{n}=n+m$) with
\begin{equation}\label{S}
S (\lambda ) = \lambda ^{2} \, (u-\lambda )^{2} - \left(\frac{N}{L} \lambda -
\frac{N-n}{L}\, u  \right) \left( \frac{M}{L} \lambda -
\frac{M-m}{L}\, u \right) \, .
\end{equation}

We have proven the following
\begin{proposition}\label{spec2}
The spectrum of $H$ given by (\ref{Hmf}) is the set $\{ \lambda _1, \lambda
_2, \lambda _{+,+}, \lambda _{+,-}, \lambda _{-,+},  \lambda _{-,-}\}$ where
$\lambda _1 =0$, $\lambda _2 = u$ and $\lambda _{\pm, \pm}$, the four
solutions of the equation $S (\lambda )=0$, have multiplicities 
$L-N_{n}-2$, $N_{n}-2$ and $1$, respectively. 
\end{proposition}

\begin{remark}
Although the eigenvalues $\lambda _{\pm ,\pm}$ can be evaluated analytically,
their expressions are too large to be useful. Numerical computation indicates
that for any values of parameters $\rho := N/L$, $\eta _{\calA }:= n/L$ and
$\eta _{\calB }:= m/L$ in the interval $(0,1/2)$, we have the
following ordering
\begin{equation}\label{order}
\begin{array}{lll}
\lambda _{-,-} < \lambda _{1} < \lambda _{+,-} < \lambda _{-,+} < \lambda
_{2}< \lambda _{+,+} & {\rm if} & u>0 \\
\lambda _{-,-} < \lambda _{2} < \lambda _{+,-} < \lambda _{-,+} < \lambda
_{1}< \lambda _{+,+} & {\rm if} & u<0 
\end{array}
\end{equation}
\end{remark}

The solutions of $S (\lambda )=0$ can be easily found if two roots are equal
to $0$ and $u$. This equation can be easily
solved also when the second term of
the right hand side of (\ref{S}) is a square of a linear
function of $\lambda $ times a constant. We shall consider three choices of
parameters having these properties:

\begin{example}
\label{example}
\begin{enumerate}
\item \label{ex0} If $\rho \le 1/2$ ($N\le M$), $\eta _{\calA }= \rho =
N_{n}/L$ ($n=N$) and $\eta _{\calB }= 0$ ($m=0$), the characteristic
polynomial  
\[
S (\lambda )= \lambda \, (\lambda -u) \left(\lambda \, (\lambda -u) +\rho (1-
\rho ) \right) 
\]
has the following roots: $\lambda _{+,-}= \lambda _{1}=0$, $\lambda _{-,+}=
\lambda _{2}=u$, $\lambda _{-,-}= \lambda _{-}$ and $\lambda _{+,+} =\lambda
_{+}$ where
\begin{equation}\label{eigen0}
\lambda _{\pm} = \frac{u}{2}\pm \sqrt{\left(\frac{u}{2} \right)^{2}+\rho
(1-\rho )} 
\end{equation}

\item \label{ex1} If $\rho =1/2$ ($N=M=L/2$) and $\eta _{\calA }=\eta _{\calB
}\equiv \eta = N_{n}/ (2L)$ ($n=m$), we have 
\[
S (\lambda )= \lambda ^{2} (\lambda -u)^{2} - \left(\lambda /2 - (1/2- \eta )u
\right)^{2} 
\]
which gives
\begin{equation}\label{eigen1}
\begin{array}{lll}
\lambda _{+,\pm} & = & \left(u+ 1/2 \pm \Delta _{+} \right) /2 \\
\lambda _{-,\pm} & = & \left(u- 1/2 \pm \Delta _{-} \right) /2 
\end{array}
\end{equation}
where
\begin{equation}\label{Delta+}
\Delta _{\pm} (\eta) = \sqrt{(u\mp 1/2)^{2} \pm 4\, u \, \eta  }
\end{equation}

\item \label{ex2} If $\rho = N/L $, $\eta _{\calA }= \rho /2$ ($n= N/2$) 
and $\eta _{\calB } = (1-\rho )/2$ ($m=M/2$), we have 
\[
S (\lambda )= \lambda ^{2} (\lambda -u)^{2} - \rho (1- \rho )\left(\lambda -
u/2\right)^{2}
\] 
which has the following roots
\begin{equation}\label{eigen2}
\lambda _{\pm,\pm}  = \frac{ u \pm \Theta _{\pm}}{2}  
\end{equation}
where $\Theta _{\pm}=\Theta _{\pm} (\rho )$ is given by
\begin{equation}\label{Theta+}
\Theta _{\pm} ^{2} = u^{2} + 2\rho (1-\rho ) \pm 2 \sqrt{\rho ^{2} (1-\rho
)^{2}+ u^{2}\rho (1-\rho )  }
\end{equation}
(Note that here $N_{n}= n+m = L/2$)
\end{enumerate}
\end{example}

For comparison with the variational method
used by Lieb--Kennedy and with the results stated in Section
\ref{groundstates}, we shall compute the ground state energy with these 
eigenvalues. Our analysis here, opposed to the previous sections, will be brief
and very selective.

\bigskip

{\it Let $N, M, n$ and $m$ be as in the Example \ref{example}.\ref{ex0}}. In
this case we have
$\lambda _{-} \le \lambda _{1} \le \lambda _{2} \le \lambda _{+}$ if $u>0$ and
$\lambda _{-} \le \lambda _{2} \le \lambda _{1} \le \lambda _{+}$ if $u<0$. 

Proceeding as in the Section \ref{groundstates}, the ground state
energy $E^{(2)}$ at $(N_{n}, N_{e})$ is given by 
\begin{description}
\item [I.] For $u>0$, 

\begin{itemize}
\item [a.] if $\calN =N_{n}+N_{e} \le L$, we have
\begin{equation}\label{gsa0}
\begin{array}{lll}
E^{(2)} (N_{n}, N_{e}) & = & \lambda _{-} (\rho ) + (N_{e}-1) \lambda _{1
} \\ 
 & = & \lambda _{-} (\rho )
\end{array}
\end{equation}
where $\rho = N/L  = N_{n}/L$;

\item [b.] if $\calN > L$, we have
\begin{equation}\label{gsb0}
\begin{array}{lll}
E^{(2)} (N_{n}, N_{e}) & = & \lambda _{-} (\rho )  + (N_{e} - L + N_{n})
\, \lambda _{2} \\ 
 & = & \lambda _{-} (\rho ) + (\calN -L) \, u
\end{array}
\end{equation}

\end{itemize}

Now, the ground state $E^{(2)}$ will be minimized with respect to $(N_{n},
N_{e})$ with $\calN =N_{n}+N_{e}$ fixed. Note that the eigenvalue $\lambda
_{-}$, given by (\ref{eigen0}), is a monotone decreasing function of $\rho $
in $0 \le \rho \le 1/2$ and a monotone increasing function in $1/2 \le \rho
\le 1$ for all $u\in \R $. This gives  

\begin{itemize}

\item [c.] if $\calN \le L $,
\begin{equation}\label{gse01}
E^{(1)} (\calN ) = \lambda _{-} (\rho ^{\ast })
\end{equation}
where $\rho ^{\ast }= \min \left((\calN -1)/L, 1/2 \right)$;

\item [d.] if $\calN > L$
\begin{equation}\label{gse02}
E^{(1)} (\calN ) = \lambda _{-} ({\hat \rho }) + (\calN -L) \, u   
\end{equation}
where ${\hat \rho }= \max \left(1/2, \calN /L -1 \right)$. 
\end{itemize}

The minimum is attained at $(N_{n}, N_{e}) = (\calN -1, 1)$ if $\calN 
< L/2$,  $(N_{n}, N_{e}) = (L/2, \calN - L/2)$ if $L/2 \le\calN 
\le 3L/2$  and at $(N_{n}, N_{e}) = (\calN -L, L)$ if $\calN > 3L/2$. As the
number of sites at the sub--lattice $\calA $ coincide with the number of
nucleus, $|\calA | = N = N_{n}$, the minimum value of 
$E^{(2)}$ is attained at the partition of $\Lambda $ into two evenly divided 
sub--lattices, $|\calA |=|\calB |$, for  $L/2 \le\calN \le 3L/2$. We recall that
the sub--lattice $\calA $ is completely occupied by nuclei and the
sub--lattice $\calB $ is empty in this example. 

\item [II.] For $u<0$, we have

\begin{itemize}
\item [a.] if $N_{e}\le N_{n}$, 
\begin{equation}\label{gsaa0}
E^{(2)} (N_{n}, N_{e})  =  \lambda _{-} (\rho ) + (N_{e}-1) \, u
\end{equation}

\item [b.] if $N_{e} >  N_{n}$, 
\begin{equation}\label{gsbb0}
E^{(2)} (N_{n}, N_{e})  = \lambda _{-} (\rho ) + (N_{n}-1) \, u
\end{equation}

\end{itemize}

Since the  minimum of  $E^{(2)}$ occurs at $N_{n}=N_{e}= \calN /2$, this gives 
\begin{equation}\label{gse00}
E^{(1)} (\calN ) = - \lambda _{+} (\calN /(2L)) + \calN /2 \, u
\end{equation}

\end{description}
$\QED $

\bigskip

{\it Let $N, M, n$ and $m$ be as in the Example
\ref{example}.\ref{ex1}}. Using the fact that, for $u>0$,  
\begin{equation}\label{D+-} 
|u-1/2| \le \Delta _{\pm} \le u+1/2 
\end{equation}
and Proposition \ref{spec2}, we have  $\mu _{1}= \lambda _{-,-}$,
$\mu _{2}= 
\cdots = \mu _{L-N_{n}-1} = 0$, $\mu _{L-N_{n}} = \lambda _{+,-}$, $\mu
_{L-N_{n}+1} = \lambda _{-,+}$, $\mu _{L-N_{n}+2}=
\cdots = \mu _{L-1}=u$ and $\mu _{L }= \lambda _{+,+}$. 

We also have 
\[
|u+1/2| \le \Delta _{\pm} \le 1/2 - u \; ,
\]
if $u<0$, which gives $\mu _{1}= \lambda _{-,-}$, $\mu _{2}=
\cdots = \mu _{N_{n}-1} = u$, $\mu _{N_{n}} = \lambda _{+,-}$, $\mu
_{N_{n}+1} = \lambda _{-,+}$, $\mu _{N_{n}+2}=
\cdots = \mu _{L-1}=0$ and $\mu _{L }= \lambda _{+,+}$. 

Observe that $\lambda _{+,-} \le  \lambda _{-,+}$ for any $u\in \R $ and $0
\le \eta \le 1/2$ because
\[
\Delta _{+}+ \Delta _{-}\ge \left\{ 2 (u^{2}+1/4)+2 |u^{2} -1/4|
\right\}^{1/2}\ge 1 \, .
\]
   

Proceeding as in the Section \ref{groundstates}, the ground state energy
$E^{(2)}$ at $(N_{n}, N_{e})$ is given by

\begin{description}
\item [I.] For $u>0$, we have
\begin{itemize}
\item [a.] if $\calN =N_{n}+N_{e} \le L$, with $N_{e}\not = L$,
\begin{equation}\label{gsa}
\begin{array}{lll}
E^{(2)} (N_{n}, N_{e}) & = & \lambda _{-,-} (\eta ) + (N_{e}-1) \lambda _{1 }
\\ 
 & = & ( u-1/2 - \Delta _{-} (\eta))/2
\end{array}
\end{equation}
(recall $\eta \equiv N_{n}/ (2L)$);

\item [b.] 
\begin{equation}\label{gsb}
E^{(2)} (0, L) = \lambda _{-,-} (0) + \lambda _{+,-} (0) = - \frac{1}{2} \; ;
\end{equation}		

\item [c.] if $\calN =N_{n}+N_{e} > L$,
\begin{equation}\label{gsc}
\begin{array}{lll}
E^{(2)} (N_{n}, N_{e}) & = & \lambda _{-,-} (\eta ) + \lambda _{+,-} (\eta ) +
\lambda _{-,+} (\eta ) + (N_{e} - L + N_{n} -1) \, u \\
 & = & ( u-1/2 - \Delta _{+} (\eta))/2 + (\calN -L) \, u
\end{array}
\end{equation}

\end{itemize}


We note, from definition (\ref{Delta+}), that $\Delta
_{\sigma } $ is a {\it monotone increasing function} of
$\eta $ if $u>0$ and $\sigma =+$ ( {\it or} if $u<0$ and $\sigma =-$). On the
other hand, $\Delta
_{\sigma } $ is a {\it monotone decreasing function} of
$\eta $ if $u>0$ and $\sigma =-$ ({\it or} if $u<0$ and $\sigma
=+$). Minimizing $E^{(2)}$ with respect to $(N_{n},N_{e})$ with $\calN
=N_{n}+N_{e}$ fixed, 
gives   
\begin{itemize}

\item [d.] 
\begin{equation}\label{gse1}
E^{(1)} (\calN ) = ( u-1/2 - \Delta _{-} (0))/2 = -1/2 
\end{equation}
if $\calN \le L $ and

\item [e.]
\begin{equation}\label{gse1'}
\begin{array}{lll}
E^{(1)} (\calN ) & = & ( u-1/2 - \Delta _{+} (1/2))/2 + (\calN -L) \, u \\
 & = &  -1/2 + (\calN -L) \, u 
\end{array}  
\end{equation}
if $\calN > L$. 
\end{itemize}

Note that the minimum is attained at $(N_{n}, N_{e}) = (0,
\calN )$ if $\calN 
\le L$ and at $(N_{n}, N_{e}) = (L, \calN -L)$ if $\calN > L$. In other words,
the ground state has no nucleus  
occupying the lattice sites of $\Lambda $ if $\calN \le L$ and $\Lambda $ is fully
occupied by nuclei if $\calN >L$. 

\item [II.] For $u<0$, we have

\begin{itemize}
\item [a.] if $N_{e}\le N_{n}$, 
\begin{equation}\label{gsaa}
\begin{array}{lll}
E^{(2)} (N_{n}, N_{e}) & = & \lambda _{-,-} (\eta ) + (N_{e}-1) \, u\\
 & = &  -(u+1/2 + \Delta _{-} (\eta))/2 + (\calN - N_{n}) \, u
\end{array}
\end{equation}

\item [b.] if $N_{e} >  N_{n}$, 
\begin{equation}\label{gsbb}
\begin{array}{lll}
E^{(2)} (N_{n}, N_{e}) & = & \lambda _{-,-} (\eta ) + (N_{n}-2) \, u + \lambda
_{+,-} (\eta ) + \lambda _{-,+} (\eta )  \\
 & = & - (u + 1/2 + \Delta _{+} (\eta))/2 + N_{n} \, u
\end{array}
\end{equation}

\end{itemize}

To minimize $E^{(2)}$ in this case one has to deal with two
contributions to the ground state energy. Note in the equations (\ref{gsaa}) 
and (\ref{gsbb}) that there is one term depending on $\eta =
N_{n}/ (2L)$ and other depending on $N_{n}$. As $N_{n}$ varies, these two
contributions moves 
the ground state energy into opposite directions. Despite this, we claim that,
if $L$ is large enough, the second contribution always dominates and the
following description holds (see Section \ref{groundstates} for similar
analysis): 
\begin{itemize}
\item [c.] 
\begin{equation}\label{gse2}
E^{(1)} (\calN ) = - \frac{1}{4} - \frac{1}{2} \; \max (\, \Delta _{+}
(\zeta),\Delta _{-} (\zeta ) \,)  + \frac{(\calN -1) }{2} \, u
\end{equation}
where $\zeta =\calN /(4L)$. Note that the minimum is attained at the neutral
point $N_{n}=N_{e}=\calN 
/2$. In addition, note that $\Delta _{+} (\zeta )=\Delta _{-} (\zeta )$  at
$\zeta =1/4$ ($\calN =L$). Since $\Delta _{\sigma } (\zeta )$ increases or
decreases according to whether $\sigma $ is positive or  negative, 
\begin{equation}\label{gse3}
E^{(1)} (\calN ) =
- \dfrac{1}{4} - \dfrac{1}{2}  \Delta _{\sigma } (\zeta )  + \dfrac{(\calN -1)
  }{2} \, u   
\end{equation}
with $\sigma =-$ if $\zeta \le 1/4 $ and $\sigma =+$ if  $\zeta >1/4$.

\end{itemize}

\end{description}
$\QED $

\bigskip

{\it Let $N, M, n$ and $m$ be as in the Example \ref{example}.\ref{ex2}}. Note
that, for all $u\in \R $ different from zero, $\Theta  
_{\sigma }(\rho )$, given by (\ref{Theta+}), is a monotone function of 
$\rho $, $0\le \rho \le 1/2$, which increases if $\sigma =+$ and decreases if
$\sigma = -$. This follows from the fact that $\Theta _{\pm}>0$ and 
\[
\Theta ^{\prime}_{\pm} (\rho ) = \frac {1-2\rho }{\Theta _{\pm}}\left[ 1\pm
\frac{1}{2\theta } \left(2 (1-2\rho )+u^{2} \right) \right] 
\]
where $\theta = \left(\rho ^{2} (1-\rho )^{2}+u^{2} \rho (1-\rho )
\right)^{1/2}$. Note, in addition, that $\Theta _{\pm} (0)= u$ and $\Theta
_{\pm} ^{\prime} (0)=\pm \infty$.

This behavior of $\Theta _{\pm}$ allows us to deduce the
expressions of  
the ground state energies $E^{(1)}$ and $E^{(2)}$ from those already
obtained for the Example \ref{ex1} by making the following replacement: 
\begin{equation}\label{r}
\begin{array}{ccc}
\pm 1/2 - \Delta _{\pm} & \longleftrightarrow & - \Theta _{\pm} \\
\pm 1/2 + \Delta _{\pm} & \longleftrightarrow &  \Theta _{\pm}
\end{array}
\end{equation}
For the item {\bf II}.c , e.g., 
\begin{equation}\label{gse4}
E^{(1)} (\calN ) = - \frac{1}{2} \; \max \left(\Theta _{+} (\rho ), \Theta
_{-} (\rho )  \right) + \frac{\calN -1}{2} \, u
\end{equation}
holds for all $u<0$, where $\rho = N /
(2L)$. We recall that the minimum of $E^{(2)}$ is attained at
$N_{n}=N_{e}$. Since the nuclei number $N_{n}=
L/2$, we conclude that $\calN =N_{n} +N_{e} = L$ is the only accessible
point in this example.  
$\QED $


\section{Conclusions}\label{conclusions}

\zerarcounters 

The nature of the ground state of the Falicov--Kimball
model with long 
range hopping matrices (mean--field matrices) was investigated. The eigenvalue
problem of this model has been entirely solved
allowing us to obtain the profile of the ground state energy. 
We have investigated the ground state energy $E^{(2)}$ as a
function of the number of nuclei and electrons $(N_{n}, N_{e})$, and $E^{(1)}$
as a function of the total number of particles 
$\calN = N_{n}+ N_{e}$,  for all values of the interaction $U$. 
This analysis revels general features of the Falicov--Kimball model which may
be conjecture to be true irrespective to whether bipartition and half--filled
band hold. The aim of this section is to point out some of these features.

Kennedy--Lieb's results on the Falicov--Kimball model (items {\bf A}-{\bf D}
in the introduction)
require the lattice to be bipartite. Let us compute, for comparison, equation
(\ref{E12}) for the bipartite mean--field model. In this case,
we have  $T^{2}= \left( {M}/{L^{2}}\right) \, \UM _{N} \oplus
\left({N}/{L^{2}} \right) \, \UM _{M}$ which gives (with $U = t\,u$ and
$2t=1$) 
\[
T^{2} + U^{2} = \left(\frac{u^{2}}{4} \, I_{N} + \frac{M}{L^{2}} \, \UM _{N}
\right) \oplus  \left(\frac{u^{2}}{4} \, I_{M} + \frac{N}{L^{2}} \, \UM _{M}
\right)  
\]
and, by using Lemma \ref{prop}, 
\[
\left( T^{2} + U^{2}\right)^{1/2} = \left( \frac{|u|}{2} \, I_{N}+
\frac{\Gamma  }{N}  \,  \UM _{N}\right)  \oplus  \left(\frac{|u|}{2} \, I_{M}+ 
\frac{\Gamma  }{M} \,  \UM _{M} \right)  
\]
where 
\beq 
\label{Gamma}
\Gamma  = \sqrt{\left(\frac{u}{2} \right)^{2}+ \frac{NM}{L^{2}}} -
\frac{|u|}{2} 
\eeq 
is equal to  
$\lambda _{+}$ if $u<0$ or $- \lambda _{-}$ if $u>0$ with
$\lambda _{\pm}$ as given in (\ref{eigen0}).

From this, we have ${1}/{2} \, {\rm Tr} \, \left(T^{2}+U^{2} \right)^{1/2} =
{|u|L}/{4} + \Gamma  $  which, when substituting in equation (\ref{E12})
($\times 
2$ in view of $t$), reproduces the ground state energy $E^{(1)}$ of the
Example \ref{example}.\ref{ex0} for $\calN \ge L$ if $u>0$ and $\calN \le
2|\calA |$ if $u<0$ (see equations (\ref{gse02}) and (\ref{gse00})). Note that
the equation (\ref{E12}) turns out to be equality in this example for $0\le
\calN \le 2L $ if $u<0$ and for $|\calN /L -1| \le 1/2$ if $u>0$.

Worth of mentioning here is the monotone behavior of $E^{(1)}$ as a function
of $\calN $. Monotonicity has been
holden generically in all examples considered (satisfying bipartition or not)
with the only exception being 
the Example \ref{example}.\ref{ex0} for $u>0$ and $0\le \calN \le L/2$ (see
equation (\ref{gse01})). This
exception, as we believe, is related to the fact that
the number of nuclei is related to the size of the sub--lattice: $|\calA
|=N_{n}$ in this example. 

For $u>0$ there is an interval of values of $\calN $, extending up to the
half--filled band point $\calN =L$, where $E^{(1)}$ remains 
constant (see equations (\ref{GSIa}), (\ref{gse01}) and (\ref{gse1})). We
believe that this behavior is generically true and is related to the
existence of first kind gap at $\calN =L$ for $u>0$. As $\calN =N_{n}+ N_{e}$
varies in 
this interval the minimum of $E^{(2)}$ is attained with the nuclei number
$N_{n}$ kept fixed. At $\calN = L$ all eigenvalues $\mu \le 0$ ($\mu \le 1/L$
for mean field hopping matrix) have been filled with electrons. As a
consequence, if  
there is a second kind gap in the spectrum of $H= T+ u \, {\rm diag} (\{\sigma
_{x}  \})$, as usually do happen for typical hopping matrices $T$, the ground
state energy density given by (\ref{eL}) develops a kink at $\rho =\calN /L
=1/2$ (see equation
(\ref{e>})) giving rise a discontinuity in the chemical potential.

We finally discuss why there is no first kind gap if $u<0$ for the mean field
models considered here. The key equation (\ref{mumu1}) shows $\mu _{1} \le
\cdots \le \mu _{N_{n}} \le u <0$ and $\mu _{j}\ge 0$ if $j>N_{n}$. So, for
any number of particles  $\calN =N_{n}+N_{e}$, the
minimum of energy is attained when all negative eigenstates are filled with
electrons: $N_{n}=N_{e}$. The sum of these eigenvalues thus gives a smooth
bulk (order $L$) contribution to $E^{(1)}$. Equations (\ref{GSII}),
(\ref{gse00}), 
(\ref{gse3}) and (\ref{gse4}), show that $e (\rho )=\dlim_{L\to \infty
}E^{(1)} (L\, \rho )/L$ is a smooth function of $\rho $, $0\le \rho \le
1$ implying absence of first kind gap. However, if sub--dominant terms of
$E^{(1)}$ are taking into account in equations
(\ref{GSII}) and (\ref{gse3}), $\dlim _{L\to \infty} \left(E^{(1)}
(L\,\rho )- L\, e (\rho ) \right)$ is continuous but non--differentiable
function of $\rho $ at $\rho =1/2$. This fact is the scar of a gap sweeped out
by the long range hopping. Opposed to the result by Kennedy--Lieb, the mean
field Falicov--Kimball model presents a phase transition at $u=0$. Observe
that there is no contradiction with item {\bf D} of the introduction since the
elements of the mean field hopping matrix are non--uniform in $L$.


\appendix
\section{Matrices of the Form $\alpha I + \beta \UM$}
\label{prop1}

\zerarcounters

\noindent
{\it Proof of Lemma \ref{prop}.} Clearly, the collection of matrices $\calC$
form a vector space. This collection also forms a commutative algebra with the
product of two matrices $C=
\alpha I + \beta \UM$ and  $D=\gamma  I + \delta \UM$ given by
\beq
\label{prod}
CD = DC = \alpha \gamma \, I + (\alpha \delta + \beta \gamma + \beta \delta R)
\UM \, .
\eeq

From equation (\ref{prod}), $D$ is the inverse of $C$, $D= C^{-1}$,
if
\beq
\alpha \gamma = 1 \;\;\;\;\;\;\;\; {\rm and} \;\;\;\;\;\;\; \alpha \delta +
\beta \gamma + \beta \delta R 
\eeq
hold. Solving these two equations we get the coefficients of $I$ and $\UM$ of the
item {\it 2}. 

To prove item {\it 3} we notice that matrices of the form $\alpha I + \beta
\UM$ are circulant matrices  
\beq
{\tilde C} = {\rm circ}(c_1, c_2, \dots , c_R) = \left(
\ba{ccccc}
c_1 & c_2 & c_3 & \ldots & c_R \\
c_R & c_1 & c_2 & \ldots & c_{R-1} \\
c_{R-1} & c_R & c_1 & \ldots & c_{R-2} \\
\vdots & \vdots & \vdots & \ddots & \vdots \\
c_2 & c_3 & c_4 & \ldots & c_1 \\
\ea
\right)
\eeq
with 
\beq
\label{c}
c_1 = \alpha + \beta \;\;\;\;\; {\rm and} \;\;\;\;\; c_2 = c_3 = \cdots = c_R=
\beta \, . 
\eeq
Circulant matrices can always be diagonalized by Fourier matrices $F$ (see
e.g. \cite{D}):
\beq
\label{sim}
\Lambda = F^{-1} \, {\tilde C} \, F
\eeq
where $\Lambda = {\rm diag} \{\lambda _1 , \lambda _2 , \dots , \lambda _R \}$
with
\beq
\label{lambda}
\lambda _j = c_1 + c_2 \omega _R^{j-1} + \cdots + c_R \omega _R^{(R-1)(j-1)}
\, .
\eeq

Substituting (\ref{c}) in (\ref{lambda}) we get
$$
\lambda _1 = \alpha + \beta \, R \;\;\;\;\; {\rm and} \;\;\;\;\; \lambda _2
= \cdots =\lambda _R = \alpha
$$ 
in view of the fact that
\beq
\label{0}
1 + \omega _R^{j-1}+ \omega _R^{2(j-1)} + \cdots + \omega _R^{(R-1)(j-1)}=0
\eeq
for all $j=2, \dots , R$.

For completeness, let us verify that (\ref{sim}) is true for matrices of the
form $C=\alpha I + \beta \UM$. Here we 
need only to Fourier transform the matrix $\UM$. Writing $F$ as the matrix
$[{\bf 1} \,{\bf u}_{R,1} \, \cdots \, {\bf u}_{R,R-1}]$ with the
vectors defined in Lemma \ref{prop} as columns and using (\ref{0}) give
$$
F^{-1} \, \UM \, F = {\rm diag} (R, 0, \dots 0)
$$ 
from which the eigenvalues of $C$ can be read.

Finally, item {\it 4} follows from item {\it 3} and this completes the proof
of the proposition. $\QED$



\end{document}